# Optical guiding in 50-meter-scale air waveguides

A. Goffin[1,2,†], I. Larkin[1,3,†], A. Tartaro[1,3], and H.M. Milchberg[1,2,3] [*]

[1]Institute for Research in Electronics and Applied Physics, [2]Dept. of Electrical and Computer Engineering, and
[3]Dept. of Physics, University of Maryland, College Park, MD 20742, USA

**Abstract:** The distant projection of high peak and average power laser beams in the atmosphere is a longstanding goal with a wide range of applications. Our early proof-of-principle experiments [Phys. Rev. X **4**, 011027 (2014)] presented one solution to this problem, employing the energy deposition of femtosecond filaments in air to sculpt millisecond lifetime sub-meter length air waveguides. Here, we demonstrate air waveguiding at the 50 meter scale, 60 × longer, making many practical applications now possible. We employ a new method for filament energy deposition: multi-filamentation of Laguerre-Gaussian LG$_{01}$ "donut" modes. We first investigate the detailed physics of this scheme over a shorter 8 m in-lab propagation range corresponding to ~13 Rayleigh lengths of the guided pulse. We then use these results to demonstrate optical guiding over ~45 m in the hallway adjacent to the lab, corresponding to ~70 Rayleigh lengths. Injection of a continuous wave probe beam into these waveguides demonstrates very long lifetimes of tens of milliseconds.

## I. INTRODUCTION

The filamentation of femtosecond laser pulses in transparent gaseous, liquid, and solid media has been of basic and applied interest for several decades [1,2]. In particular, filamentation in air or other gases has been demonstrated from the infrared [3,4] through the ultraviolet [5], with application to supercontinuum generation [6], THz generation [7-10], remote sensing [11], and the triggering and guiding of high voltage electrical discharges [12,13].

Filamentation in air commences when a pulse self-focuses owing to the instantaneous electronic and delayed molecular rotational nonlinearity of air constituents [14,15]. The pulse continues to self-focus until collapse is arrested by plasma defocusing from high field ionization of air molecules. In air filaments, the competition between self-focusing and ionization defocusing limits the core intensity to $I_{\text{core}} < \sim 10^{14}$ W/cm$^2$ [16], and the composite effect is a self-guided, high intensity optical filament with a core diameter $d_{\text{core}} < \sim 200$ μm surrounded by a lower intensity 'reservoir' region with which it continually exchanges energy during propagation [17]. The length of the filament, $\Delta z_{fil}$, is usually considered to be the axial extent over which the filament core remains sufficiently intense for a particular application, with $\Delta z_{fil}$ typically orders of magnitude greater than the effective Rayleigh range $\sim \pi (d_{core}/2)^2/\lambda$ corresponding to $d_{core}$, where $\lambda$ is the laser central wavelength. For collapse and filamentation of a collimated beam of diameter $d_{beam}$, the filament length is governed by the confocal parameter of the overall beam, $\Delta z_{fil} \sim 2z_0 \sim 2\pi(d_{beam}/2)^2/\lambda$ [18]. Therefore, filaments generated by centimeter diameter beams ($d_{beam} \sim 1$ cm) can be several hundred meters long; such extended distances are of high interest for applications.

While femtosecond filaments in air can deliver high peak intensity over extended distances, leading to many of the applications cited earlier, the energy in a filament core is self-limited to approximately the millijoule level: $\varepsilon \sim I_{\text{core}} d_{\text{core}}^2 \tau < \sim 1$ mJ, using the parameters from above and pulse duration $\tau \sim 100$ fs. For example, a laser with a 1 kHz repetition rate could deliver only ~1 W of average power in a single filament core. From the perspective of high average power applications, this is a severe limitation.

---
[†] These authors contributed equally to this work.
[*] milch@umd.edu

There is one aspect of filamentation, however, that can be harnessed to enable delivery of very high average laser powers over extended distances in air: its ability to imprint, via localized air heating, very long-lived optical waveguide structures [19]. This occurs as follows: The propagating femtosecond filament core generates plasma [20] and coherently excites rotational wavepackets in $N_2$ and $O_2$ molecules [21, 22]. Subsequent thermalization accompanying plasma recombination (<10 ns) and collisional decoherence of the rotations (<100 ps) [23] is much faster than the hydrodynamic response timescale of neutral air ($\sim d_{\text{core}}/2c_s \sim 300$ ns, for air sound speed $c_s \sim 300$ m/s), and this manifests as impulse heating of the air and a pressure spike extending for the full length of the filament. The pressure spike launches a single-cycle cylindrical acoustic wave which propagates away after a few microseconds, leaving a long lifetime density depression, or "density hole", on axis [24,25]. The density hole lasts several milliseconds, characteristic of the thermal diffusion timescale of air. An array of such holes can form a guiding structure for a secondary laser pulse.

In our prior work [19], we demonstrated that an array of 4 filaments, generated by the 4 lobes of a $TEM_{11}$ beam, could form an air waveguide structure in two timescale regimes. In the short timescale "acoustic" regime, lasting several hundred nanoseconds, the on-axis collision of the acoustic waves from the 4 filaments forms an increased air density waveguide core. The acoustic waves then propagate away over several microseconds leaving 4 density holes. In the long timescale "thermal" regime, lasting up to milliseconds, the density holes merge circumferentially by thermal diffusion to form an effective "moat" or cladding around the unperturbed air at the centre of the beam. The central unperturbed air forms air waveguide core. The 4-lobed $TEM_{11}$ mode was formed by an effective binary phase mask, where alternating segments impose a relative spatial phase shift of $\pi$. The resulting waveguides were 70 cm long, and guided an injected probe pulse with spot size ~150 μm.

In this paper, we demonstrate optical guiding in the longest thermal regime air waveguides generated by far, 45 m in length. From the confocal parameter scaling of filament length discussed above, this required filamentation of a beam with $d_{beam} \sim 6$ mm. At the same time, to form an effective air waveguide cladding, such a beam should form a sufficient number of filaments on its periphery to ensure circumferential coverage. To accomplish this, we took a different approach from the use of binary phase masks [19]; here we use a smooth Laguerre-Gauss $LG_{01}$ mode to initiate random filamentation in the donut ring. While a $n$-segment binary phase mask can, in principle, seed filaments at each of the $n$ beam lobes imposed by the mask, in practice it is difficult to ensure that the lobes have equal energy and locally smooth phase fronts. By contrast, we demonstrate a method to produce a high energy, high quality Laguerre-Gaussian $LG_{01}$ "donut" or "phase vortex" mode [26] so that filamentation is seeded with low level intensity or phase noise that is more uniformly distributed across the mode. Unlike with a binary phase mask, the number of filaments automatically scales with beam size provided that the local laser fluence remains constant, ensuring circumferential coverage of the generated cladding. Furthermore, an LG beam propagates as a single mode with a well-defined Rayleigh range, the distance over which its peak intensity drops by a factor of two owing to diffraction. By contrast, a beam generated by a high order binary phase mask is highly multimodal with much greater diffractive spreading. We note that in other recent work, filament arrays have been generated with amplitude or phase masks [27, 28].

## II. LG$_{01}$ BEAM REQUIREMENTS FOR AIR WAVEGUIDE GENERATION

The electric field of a LG$_{01}$ vortex mode is $E_{01}(r,\varphi,z) = \sqrt{eF_0}\,(w_0/w(z))(\sqrt{2}r/w(z))e^{-i2\tan^{-1}(z/z_0)}\,e^{-ikr^2/2R(z)}e^{-(r/w(z))^2}e^{i\varphi}$, where $F_0$ is the peak fluence (J/cm$^2$) in the mode's ring, and where $w(z) = w_0(1+(z/z_0)^2)^{1/2}$, $R(z) = z(1+(z_0/z)^2)$, and $z_0 = \pi w_0^2/\lambda$ are the mode's spot size, phasefront curvature, and Rayleigh range, respectively. The mode's peak fluence lies on a ring of diameter $d_{ring} = \sqrt{2}\,w(z)$. Because the LG$_{01}$ beams of this experiment are highly collimated (propagation range $z < z_0$), we consider $d_{ring} \cong \sqrt{2}\,w_0$ for making estimates.

An interesting aspect of vortex beam filamentation is that the beam ring self-focuses to a narrow shell before filaments nucleate, as borne out by our measurements and simulations shown later. This is in contrast to a non-vortex beam where filaments can grow throughout the beam cross section. For an LG beam, filaments are thus located on a well-defined ring, ideal for air waveguide generation. This suggests that for fixed $F_0$, the number of filaments will scale as the ring circumference ($\propto d_{ring}$). Indeed, for peak laser fluence $F_0$ in the LG$_{01}$ pulse of energy $\varepsilon_{LG}$, the number of filaments formed is

$$n_{fil} \sim 1.2\,(\varepsilon_{LG}/\varepsilon_{cr})^{1/2} = 1.8\,(F_0/\varepsilon_{cr})^{1/2}\,d_{ring} \qquad (1)$$

where $\varepsilon_{cr} \sim P_{cr}\tau$ and $P_{cr} \sim \lambda^2/2\pi n_0 n_{2,eff}$ [29] are the critical energy and power for self-focusing collapse for a small section of the LG$_{01}$ ring containing energy $\varepsilon_{cr}$, and where Eq. (1) is adapted from ref. [30]. The critical power ranges over $P_{cr} = 13\,\text{GW} - 3\,\text{GW}$ for laser pulsewidths $\tau = 45 - 300$ fs used in these experiments. This stems from the pulsewidth dependence of the effective nonlinear refractive index, $n_{2,eff} \sim 0.8 - 3.8 \times 10^{-19}$ cm$^2$/W in this range, owing to the delayed molecular rotational nonlinearity [14]. For example, for one set of our experimental parameters ($\varepsilon_{LG} = 90$ mJ, $F_0 = 0.21$ J/cm$^2$, $\tau = 100$ fs, and $d_{ring} = 0.45$ cm), the expected number of filaments is $n_{fil} \sim 20$.

Considering the air waveguide as a step index fibre with $V$ parameter [31] $V = (2\pi a/\lambda)(n_{co}^2 - n_{cl}^2)^{1/2}$ enables an estimate of the cladding air density reduction needed for guiding. Here $a = d_{ring}/2$ is the waveguide core radius and the core and cladding indices are $n_{co} = n_0 + \delta n_{co}$ and $n_{cl} = n_0 + \delta n_{cl}$, with $\delta n_{cl}/n_0 \ll 1$ and $\delta n_{co} = 0$ for a thermal waveguide. The condition for guiding of a lowest order mode is then $V \sim \sqrt{2}\pi(d_{ring}/\lambda)|\delta n_{cl}|^{1/2} > 2.405$ [18], so that the minimum air index reduction in the cladding is $|\delta n_{cl}| \sim 10^{-8}$. This extremely small index decrement is reflective of the large waveguide core size $a \sim 2$ mm, but it is unrealistic: it is significantly smaller than the index fluctuation associated with our measured lab air turbulence level of $C_n^2 = 6.4 \times 10^{-14}$ (m$^{-2/3}$) [32], which gives $|\delta n_{turb}| \sim 10^{-7}$ across the air waveguide core. If we impose the conservative condition $|\delta n_{cl}| = 10\,|\delta n_{turb}|$ so that the cladding moat depth greatly exceeds the turbulence fluctuation level, the relative air density reduction in the cladding (or refractive index contrast) should be $|\Delta N_{cl}|/N_0 = |\delta n_{cl}|/(n_0 - 1) > \sim 0.4\%$.

Under our conditions, the maximum relative depth of a filament-induced density hole after recombination and thermalization is $\Delta N_h^{max}/N_0 \sim 1 - T_0/T_h \sim 0.25$, where $T_0 \sim 300$K is ambient air temperature and $T_h \sim 400$K is the typical peak temperature of the density hole [25]. The hole depth slowly declines by thermal diffusion according to $\Delta N_h(t)/N_0 \sim (\Delta N_h^{max}/N_0)(1+4\alpha t/R_0^2)^{-1}$, where $\alpha = 0.20$ cm$^2$/s is the thermal diffusivity of air and $R_0 \sim 100$ μm is the initial density hole radius [25]. Thus, the depth of a density hole will decline from 0.25 to 0.004 over a time $t \sim 10$ ms, giving a good estimate of our thermal waveguide cladding lifetime, as we will see from later comparison with experiment.

The remaining question is whether there is adequate azimuthal cladding coverage by the thermally diffusing density holes. This is what determines the LG$_{01}$ mode energy needed for the filaments leading to cladding formation. After ~1 μs of delay, the density hole widens by thermal diffusion [25], giving $R_h \sim R_0(1+4\alpha t/R_0^2)^{1/2} \sim 1$ mm by $t = 10$ ms, so the density hole spacing $\Delta x_{fil}$ should be ~1 mm at most. For example, for $d_{ring} = 4.5$ mm, the number of filaments needed around the LG$_{01}$ ring is $n_{fil} \sim \pi d_{ring}/\Delta x_{fil} \sim 15 - 25$ for $\Delta x_{fil} \sim 0.5 - 1$ mm, which is consistent with the earlier estimate using $E_{LG} \sim 90$ mJ. This agrees quite well with the energies used in the experiments, as discussed below.

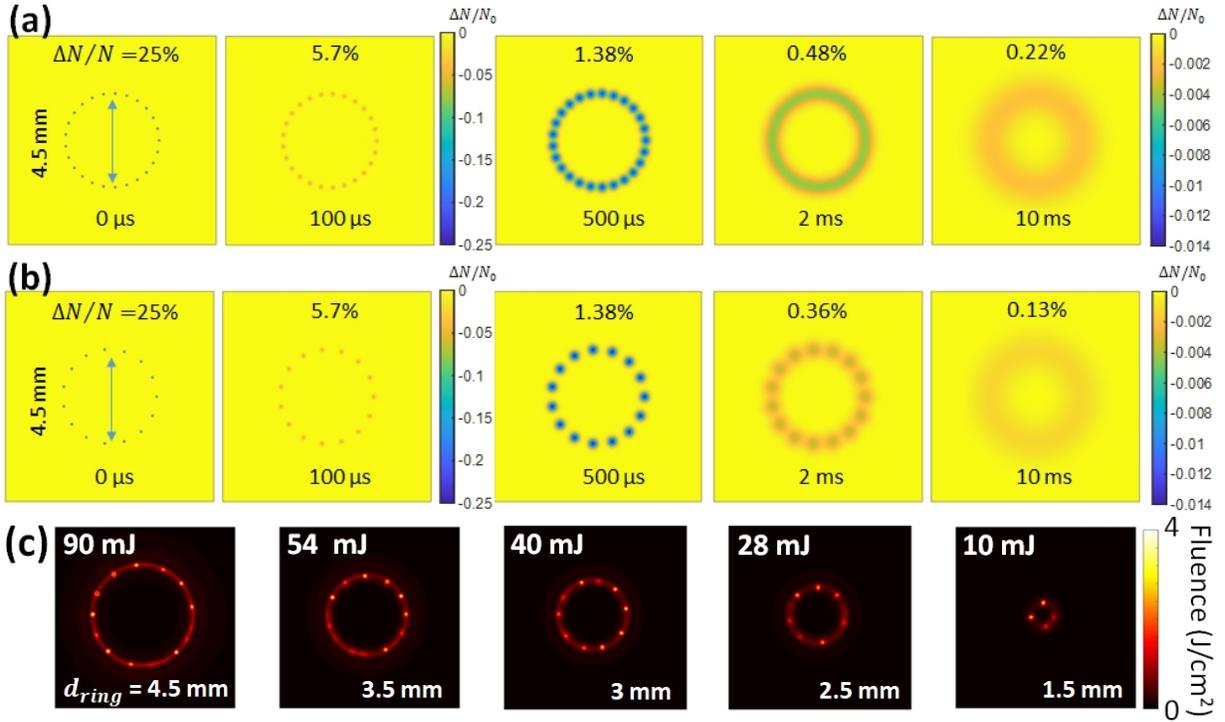

**Figure 1.** (a) Thermal response of air to a ring array of 25 filament-initiated density holes of initial depth $|\Delta N|/N = 0.25$ and peak temperature increase $\Delta T = 100$K. with increasing delay the density holes merge by thermal diffusion to form a nearly continuous cladding moat around the central unperturbed air density core. (b) same as (a), except for 15 filament-initiated density holes. (c) 3D+1 YAPPE propagation simulation (Appendix A) of the onset of filamentation of 100fs LG$_{01}$ pulses of varying diameter and energy for constant initial peak fluence $F_0 = 0.21$ J/cm$^2$. The LG$_{01}$ pulses were initialized using a white noise amplitude mask with fluence standard deviation $0.01 F_0$ to seed filament nucleation.

Figure 1(a) shows hydrodynamic simulations [19] of thermal diffusion for cases of $n_{fil} = 25$ and $n_{fil} = 15$ filament-induced density holes spread uniformly on a 4.5 mm diameter ring in air. The initial energy deposition for each filament is taken to be a Gaussian with $1/e$ radius $R_0 = 50$ μm, with temperature increase $T_h - T_0 = 100$ K, giving initial density hole depth $\Delta N_h^{max}/N_0 = 0.25$ (as discussed earlier). This matches typical filament conditions [25]. It is seen that by ~1 ms, the density holes have sufficiently merged to form a nearly continuous cladding "moat" surrounding the unperturbed air core. Even out to delays of 10 ms, $|\Delta N_{cl}|/N_0 \sim 0.22\%$ indicates that reasonable guiding confinement could be expected based on the conservative estimate made earlier in this section. As will be seen, this is borne out in our guiding experiments. Owing to the much wider waveguides produced in the current experiment, their lifetime is expected to be considerably greater than the millisecond duration of the ~200 μm wide guides of our earlier work [19].

Figure 1(c) shows a sequence of 3D+1 (3 space dimensions plus time) YAPPE simulations ([33], Appendix A) of the filamentation of 100 fs $LG_{01}$ pulses for several beam waists $w_0 = d_{ring}/\sqrt{2}$, with pulse energy scaled to maintain constant initial peak fluence $F_0$. In each case, the pulse was propagated to the onset of filamentation. The $LG_{01}$ pulses were initialized with a white noise amplitude mask with a fluence standard deviation of 1% of $F_0$ to seed filamentation. It is seen that the $LG_{01}$ ring self-focuses to a narrow shell before filaments nucleate. For each beam size, repeated simulations with different white noise masks of the same standard deviation show a similar number of filaments. For increasing $d_{ring}$ with $F_0$ constant, the simulations show that $n_{fil} \propto \sqrt{\varepsilon_{LG}}$, in agreement with ref. [30] and Eq. (1). For $d_{ring} = 4.5$ mm, $n_{fil} = 15 - 20$, in reasonable agreement with the estimate based on Eq. (1).

To conclude this section, it is worth explaining why an $LG_{0m}$ mode with $m = 1$, rather than $m > 1$, is preferred for generating a ring of filaments. Practically, a $LG_{0m}$ mode is generated by passing a Gaussian beam of spot size $w_0$ through an $m^{th}$ order spiral phase plate. For a given pulse energy $\varepsilon_{LG}$, the peak fluence in the $LG_{0m}$ ring is $F_0 = (2\varepsilon_{LG}/\pi w_0^2)g(m)$, where $g(m) = (m/e)^m/m!$ is a decreasing function of $m$: for example, $g(1)/g(5) \sim 2$. So an $LG_{01}$ mode provides the highest ring fluence $F_0$ for a given laser energy. In addition, higher order vortex modes diverge faster: $\Delta\theta_{0m} = \sqrt{m}\,\Delta\theta_{01}$, where $\Delta\theta_{0m}$ is the angular divergence of the ring in an $LG_{0m}$ mode.

## III. EXPERIMENTAL SETUP

Our air waveguide experiments were separated into medium range (< 8 m, in-lab) and longer range (< 50 m, in hallway outside lab) experiments depicted in Figs. 2(a) and 2(b). Filaments for the air waveguide were generated using a 10 Hz Ti:Sapphire laser system with $\lambda_0$= 800 nm and pulsewidth $\tau = 45 - 300$ fs (FWHM), with the pulsewidth adjusted using the grating compressor. The air-waveguided probe was a $\lambda_0$= 532 nm, τ ~7 ns, 1 mJ pulse from a frequency-doubled Nd:YAG laser. Filaments were generated with post-compression energies up to 120 mJ, controlled by a λ/2 waveplate and a thin film polarizer (TFP).

The key diagnostics for the in-lab experiments (Fig. 2(a)) were a translatable helium cell-based imaging system [34,35] and a microphone array [32]. Filaments from the air side were terminated over a < ~ 4 mm air-to-helium transition in the helium cell's slow-outflow nozzle (see Fig. 2(a)), owing to the ~20 × lower nonlinear refractive index of helium compared to air [36]. This enabled direct in-flight linear imaging, through the helium cell, of $LG_{01}$ beam filamentation and guided mode evolution along the 8 m propagation path. The microphone array, 64 synchronized microphones spanning 126 cm, was used to map axial energy deposition profiles of the beams generating filament waveguides. The array measured single shot records of the filament energy deposition per unit length over the full filament propagation path [18,32]. The local filament energy absorption drives the generation of air density holes, which form the air waveguide cladding.

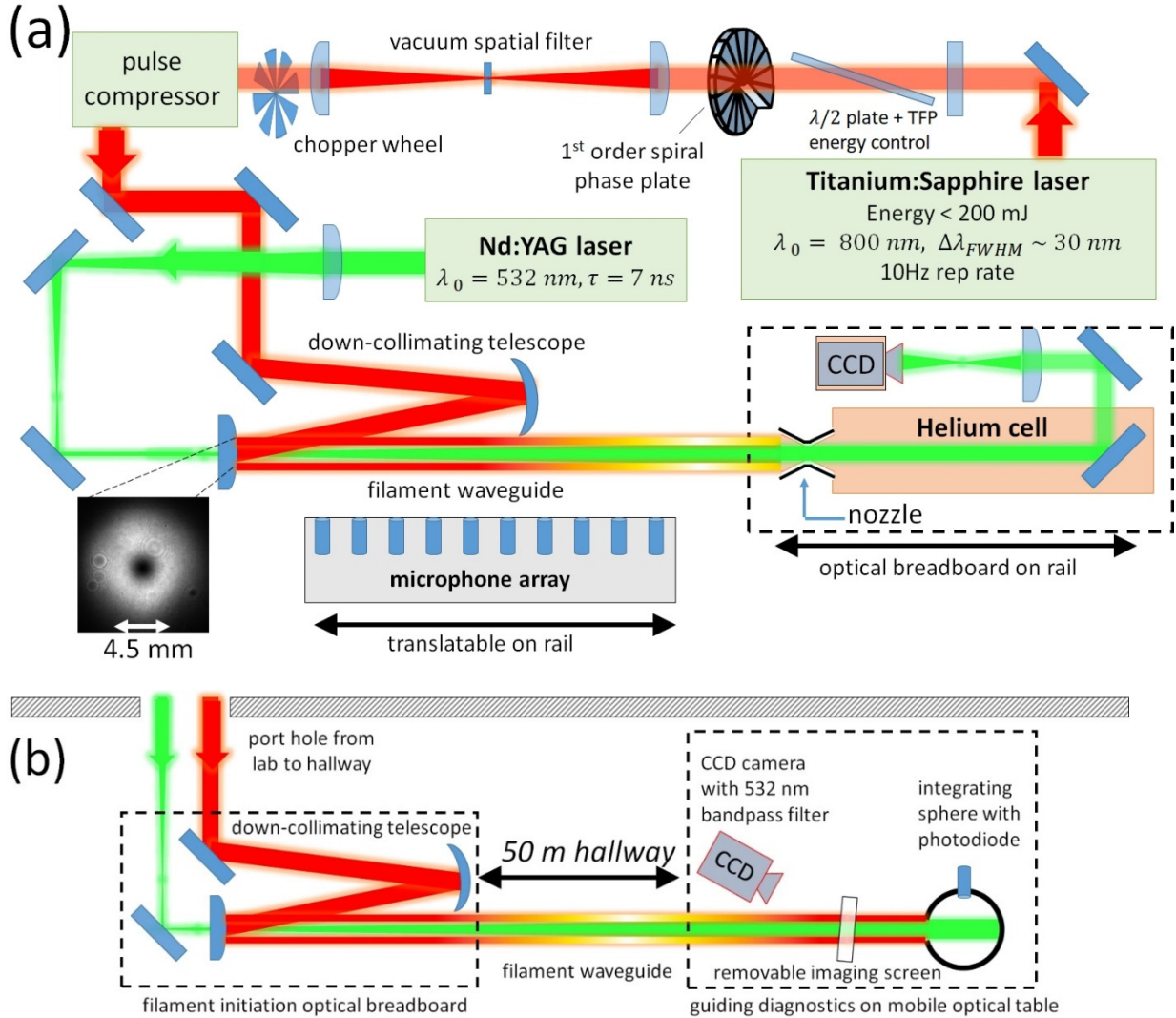

**Figure 2. (a)** Experimental setup for 8 m in-lab guiding experiments. **(b)** Setup for 50 m hallway guiding experiments. The alumina ceramic imaging screen was used for guided mode measurements for the λ=532 nm, 7 ns, 1 mJ probe pulse (Fig. 8). The integrating sphere was used in guiding lifetime measurements using a CW λ=532 nm probe beam (Fig. 9). The CCD camera and integrating sphere used 532 nm interference filters.

The high energy $LG_{01}$ donut mode used to form air waveguides is generated by passing the pre-compressor laser pulse through a 16-step first order spiral phase plate ($2\pi$ phase shift around the beam), followed by a vacuum spatial filter to smooth the beam, removing high spatial frequency nonuniformities including those introduced by the phase plate steps. The resulting beam is then passed through the pulse compressor, followed by a $3\times$ down-collimating reflective telescope (used for both the lab and hallway experiments) to produce a high quality $LG_{01}$ beam whose waist is located at the telescope output. The in-lab experiments used ring diameters of $d_{ring} = 4.5$ mm and 3 mm, giving $w_0 = d_{ring}/\sqrt{2} \simeq 3.2$ mm and 2.1 mm, where $w_0$ is the $e^{-2}$ intensity radius of the corresponding lowest order LG mode. These corresponds to Rayleigh ranges $z_0 \sim \pi w_0^2/\lambda \sim 40$ m and 17 m. For the hallway, $d_{ring} = 5.6$ mm and $z_0 \sim 60$ m. The aim was to produce "natural" LG$_{01}$ filamentation generated purely by nonlinear self-focusing and without the assistance of external focusing to maximize the beam's Rayleigh range and, therefore, the waveguide length. In both the in-lab and hallway experiments, the green probe laser pulse was injected through the second mirror of the down-collimating telescope and co-propagated with the waveguide-forming LG$_{01}$ beam. A lens is placed upstream of the telescope mirror in the probe beamline to form a telescope whose effect is a $\sim f/950$ defocus on the probe for injection into the waveguide.

## IV. EXPERIMENTAL RESULTS AND DISCUSSION

### (a) Measurements of LG$_{01}$ filament formation

Experiments for air waveguides up to 8 m took place inside the lab with the configuration shown in Fig. 2(a). To determine the optimum laser pulsewidth for extended filament generation, we first employed the microphone array to measure the pulsewidth-dependent energy deposition profiles for single filaments generated with Gaussian LG$_{00}$ pulses with $w_0 = 1$ mm ($z_0 = 4$ m). The results are shown in Fig. 3(a), where the curves are 100 shot averages of concatenated 1.26 m longitudinal sections. Here the peak pulse power was kept constant at $\sim 6P_{cr}$, accounting for pulsewidth-dependent $n_{2,eff}$ [14]. It is seen that while the peak energy deposition is highest using shorter pulses, owing to more plasma generation from optical field ionization at the highest intensities, the axial extent of filamentation increases at longer pulsewidths. This originates from the increased contribution of molecular rotation of N$_2$ and O$_2$ to $n_{2,eff}$ of air as pulsewidths exceed ~50 fs [14], leading to extended filamentary propagation [20]. Based on these results, we chose 100 fs pulses for the in-lab experiments because their filaments extended ~8 m to the end of the available lab space. Pulse energies in the range 80-90 mJ generated the needed number of filaments ($n_{fil} \sim 20$) for good azimuthal cladding coverage; filaments begin within 1 m of the down-collimating telescope. Figure 3(b) shows the composite microphone array signal from overlapping 1.26 m longitudinal sections over the propagation range of the filamenting LG$_{01}$ pulse, where it is seen that filament energy is absorbed over a range > 7 m.

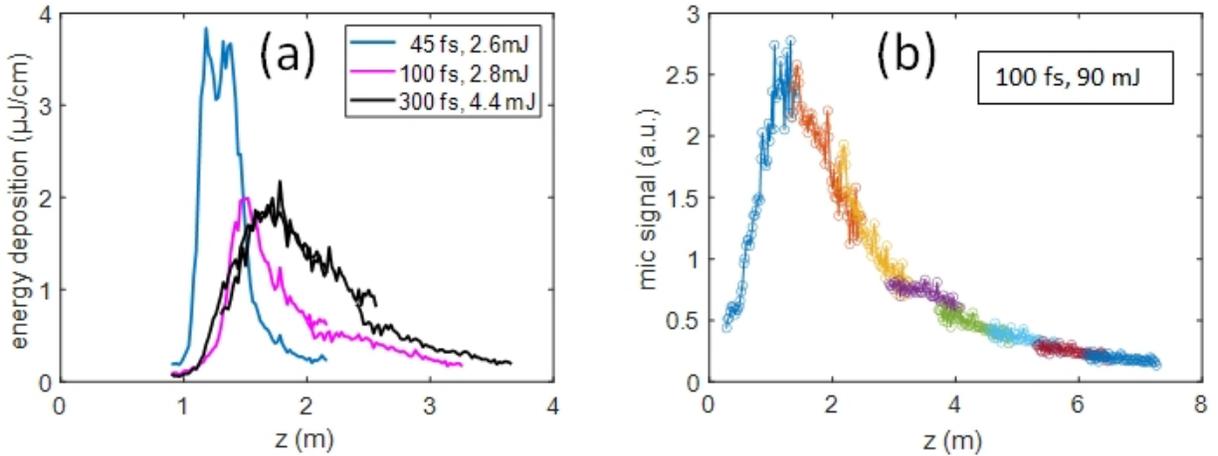

**Figure 3. (a)** Single filament energy absorbed per unit length vs. propagation distance measured by microphone array for laser pulsewidths 45 fs, 100 fs, and 300 fs. Pulse energies were chosen to keep peak power constant at $\sim 6P_{cr}$. Curves are concatenated 126cm longitudinal sections, averaged over 100 shots. The array was 3 mm from the beam axis. **(b)** Acoustic signal from filamentation of 100 fs, 90 mJ LG$_{01}$ pulse. The microphone array was 5

mm from the beam axis. Concatenated and overlapped 126cm longitudinal sections, averaged over 100 shots, are plotted in different colours.

To capture the onset and formation of filaments in a LG$_{01}$ beam, we used the in-lab translatable helium cell to interrupt propagation and directly image the beam cross-section. Figure 4(a) shows single shot beam images from a pulse energy scan at a fixed longitudinal location ($z = 1.5$ m from the down-collimating telescope), using a LG$_{01}$ beam with $d_{ring} = 3.8$ mm to highlight the pre-filamentation phase of propagation. Already with 7.2 mJ, the beam's ring has narrowed, with the clear enhancement of slight beam nonuniformities. The ring width dramatically narrows further as pulse energy increases, with filaments nucleating at the local beam intensity maxima. For a 34 mJ, 45 fs, $d_{ring} = 4.5$ mm beam, Fig. 4(b) shows single shot images of increasing multi-filamentation with propagation up to $z \sim 2.6$ m, after which the total number of filaments stays roughly constant at $n_{fil} \sim 15 - 20$, where both bright and fainter (incipient) filamentary hotspots are counted. This is consistent with the $n_{fil}$ estimates made in Sec. II and the 3D+1 simulations of Fig. 1(b). While the filament locations on the ring change shot to shot, their azimuthal distribution remains relatively uniform: this illustrates the advantage of air waveguide generation by a smooth LG$_{01}$ mode over a beam generated by a binary phase plate.

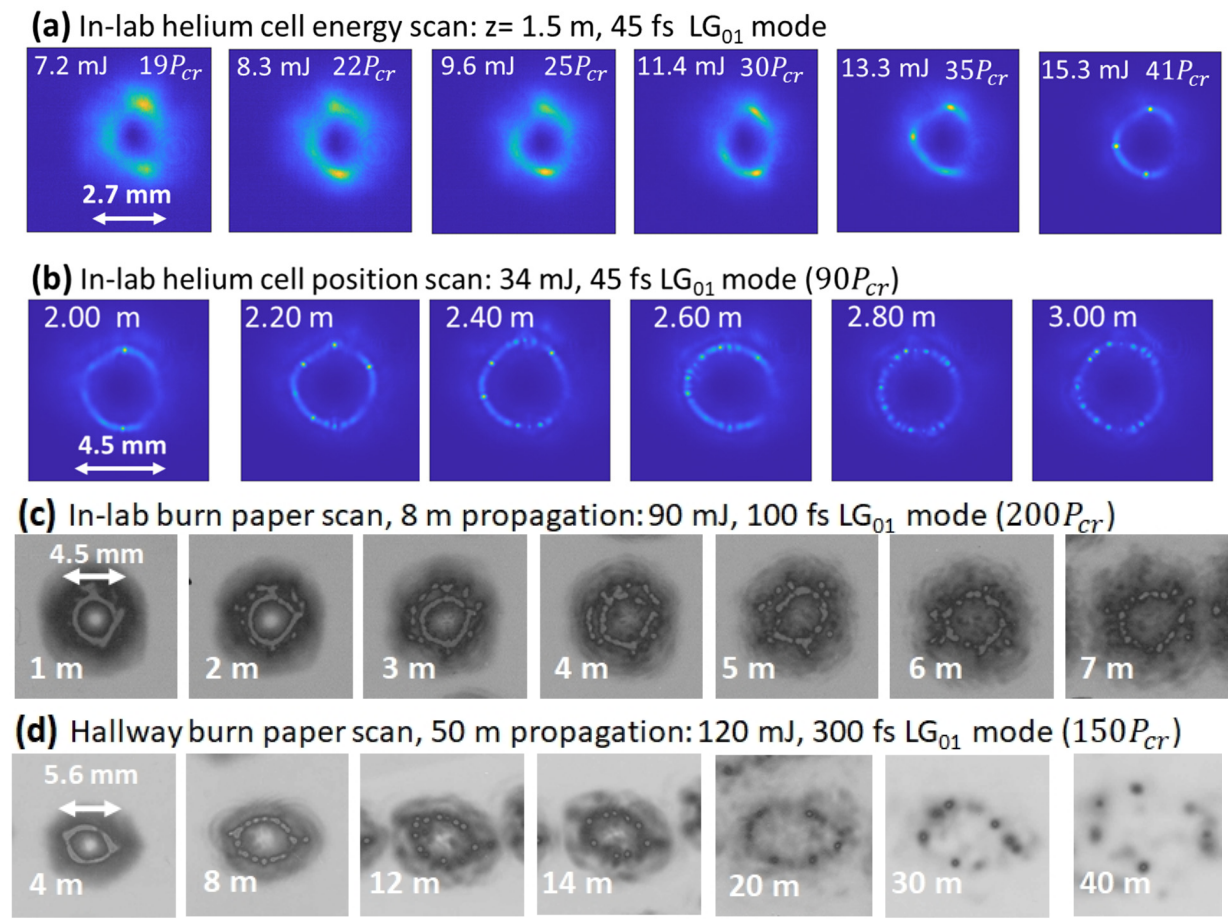

**Figure 4. (a)** Mode images from helium cell of pulse energy scan showing progressive narrowing of LG$_{01}$ ring and nucleation of filaments. The peak pulse power in units of $P_{cr}$ is shown in each image. **(b)** In-lab images from longitudinal scan of helium cell for an LG$_{01}$ pulse with $P = 90 P_{cr}$, showing nucleation of filaments and their progressive increase in number. **(c)** In-lab burn paper patterns taken for comparison with helium cell images. **(d)** Hallway burn patterns over length of propagation range.

Additional beam measurements were taken using burn paper, needed for the hallway experiments in the absence of the helium cell. To capture single shots, the paper was quickly passed through the filamenting beam as it pulsed at the 10 Hz repetition rate of our laser system. To compare with the helium cell images, an in-lab burn paper scan vs. distance is shown in Fig. 4(c). The scan qualitatively shows (despite the paper saturation) good azimuthal filament coverage over the 8 m propagation range, consistent with the ≤ 3 m helium cell measurements in Fig. 4(b). The hallway burn patterns up to 42 m similarly show good azimuthal coverage, with the number of filaments decreasing at the longest distances owing to the beam intensity decrease from diffractive spreading.

**(b) 8 m air waveguiding experiments**

Having confirmed that both the longitudinal energy deposition profile and azimuthal filament coverage of filamenting LG$_{01}$ beams are consistent with generating very long air waveguides, we now present experimental demonstration of guiding, first over 8 m in the laboratory. Injection of the λ=532 nm probe pulse into the air waveguide was delayed by 800 μs after LG$_{01}$ multi-filament initiation to ensure that the individual density holes thermally diffused into a relatively continuous lower-density cladding moat. As discussed, the λ=532 nm probe pulse was coupled into the guide in co-propagating geometry (unlike the counter-propagating geometry of ref. [19]) by passing it through a lens followed by the dielectric curved mirror of the 800 nm reflective telescope (see Fig. 2(a)), imposing a diverging phase curvature with Rayleigh length $z_{pr} \sim$ 60 cm, equivalent to defocusing at ~ $f$/950.

Use of the diverging probe provided a more rigorous test of the waveguide than a collimated probe, enabling demonstration of guiding over ~$13 z_{pr}$ in the lab and ~$70 z_{pr}$ in the hallway. For an air waveguide with an index contrast $|\Delta N_{cl}|/N_0 = |\delta n_{cl}|/(n_0 - 1) \sim$ 0.5% (see prior discussion and Fig. 1(a)) and $d_{ring} = 4.5$ mm, $V = 44$ and the numerical aperture is $NA = \lambda V/\pi d_{ring} \sim 1.7 \times 10^{-3}$, supporting coupling f-numbers $f_\# = 0.5/NA \sim$ 300 or larger. This guide will easily trap and guide our defocusing~$f$/950 probe pulse; such a guide is highly multimodal, trapping ~ $V^2/2 \sim 10^3$ modes [31].

As depicted in Fig. 2(a), the guided beam was directly imaged from the helium cell-scanned exit of the air waveguide onto a CMOS camera (through a 532 nm bandpass filter); the guided beam was effectively imaged *inside the waveguide* as function of propagation distance. Co-propagating supercontinuum light generated by the multiple LG$_{01}$ filaments was attenuated by a bandpass filter and linear polarizer in front of the camera. The camera exposure was temporally gated to eliminate any residual supercontinuum light.

Figure 5 shows optical guiding of the probe beam injected at 800 μs delay into an air waveguide formed by an 80 mJ, 100 fs LG01 pulse with $d_{ring} = 4.5$ mm, with Fig. 5(a) and (b) showing the probe maintaining a constant ~4 mm guided beam diameter over an effective Rayleigh range ~$13 z_{pr}$. As predicted, these beams are highly multimodal. For the shorter $z$ images in Fig. 5(a), one can clearly see the imprints of the circumferential array of density holes on the outside edge of the guided beam. Without the guide present, the probe rapidly diverges. The gap in measurements between $z = 5$ m and 8 m (Fig. 5(c)) is due to a gap in helium cell travel constrained by our optical table arrangement.

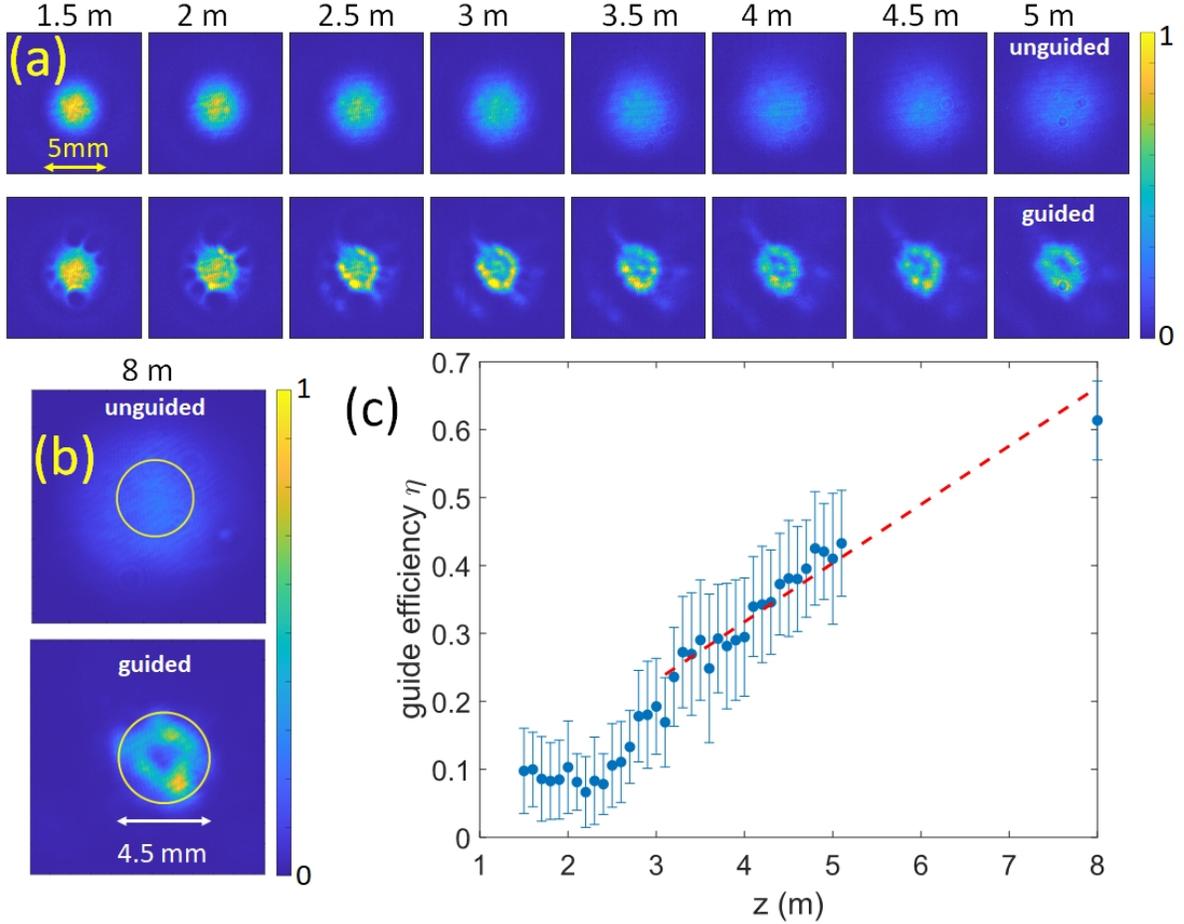

**Figure 5.** Optical guiding of a 1 mJ, 7 ns, λ=532 nm probe pulse injected at 800 μs delay into an air waveguide formed by a 90 mJ, 100fs $LG_{01}$ pulse with $d_{ring} = 4.5$ mm. **(a)** 50 shot average images of unguided (top row) and guided (bottom row) mode vs. propagation distance. The guided mode diameter is a constant ~4 mm. **(b)** Images at $z = 8$ m of unguided and guided modes. The $d_{ring} = 4.5$ mm diameter circles contain the guided and unguided energy used in the definition of guiding efficiency. **(c)** Guiding efficiency $\eta = (E_g - E_{ug})/(E_{tot} - E_{ug})$ vs. propagation distance, measured with the helium cell. Each point is a 50 shot average, with the bars showing the ± standard deviation. The gap between points at $z = 5.3$ m and $z = 8$ m is due to helium cell travel constrained by our optical table arrangement. The dashed red line is a linear fit to the points > 3m to guide the eye.

To characterize guiding efficiency, a chopper was inserted into the waveguide-generating beam (see Fig. 2(a)) so every other probe laser pulse was guided; 100 consecutive images were saved for every point in Fig. 5, and the average of 50 guided and 50 unguided laser shots is displayed. The guiding efficiency metric defined in our previous work [19] is

$$\eta = (E_g - E_{ug})/(E_{tot} - E_{ug}) , \qquad (2)$$

where $E_g$ and $E_{ug}$ are the guided and unguided energy within the central mode area (that is, with and without the waveguide), and $E_{tot}$ is the total beam energy. The efficiency $\eta$ ranges from 0 to 1 and gives the fraction of energy retained within the waveguide that would otherwise diffract from within the guided mode area (inside the circle in Fig. 5(b)).

Figure 5(c) plots the guiding efficiency $\eta$ vs. propagation distance for the waveguide. The increase in $\eta$ with distance results from its definition: as $z$ increases, the unguided mode contributes decreasingly to the beam energy within the circle defining the guided mode. The maximum guiding efficiency of ~60% is comparable to that achieved over < 1 m of guiding in our previous work [19]. Each point is a 50 shot average with the vertical bars showing the ± standard deviation. The fluctuations are dominated by shot-to-shot fluctuations in probe beam energy—the guided mode profiles are very stable, and the 50 shot average images in Fig. 5(a) closely resemble those of individual shots.

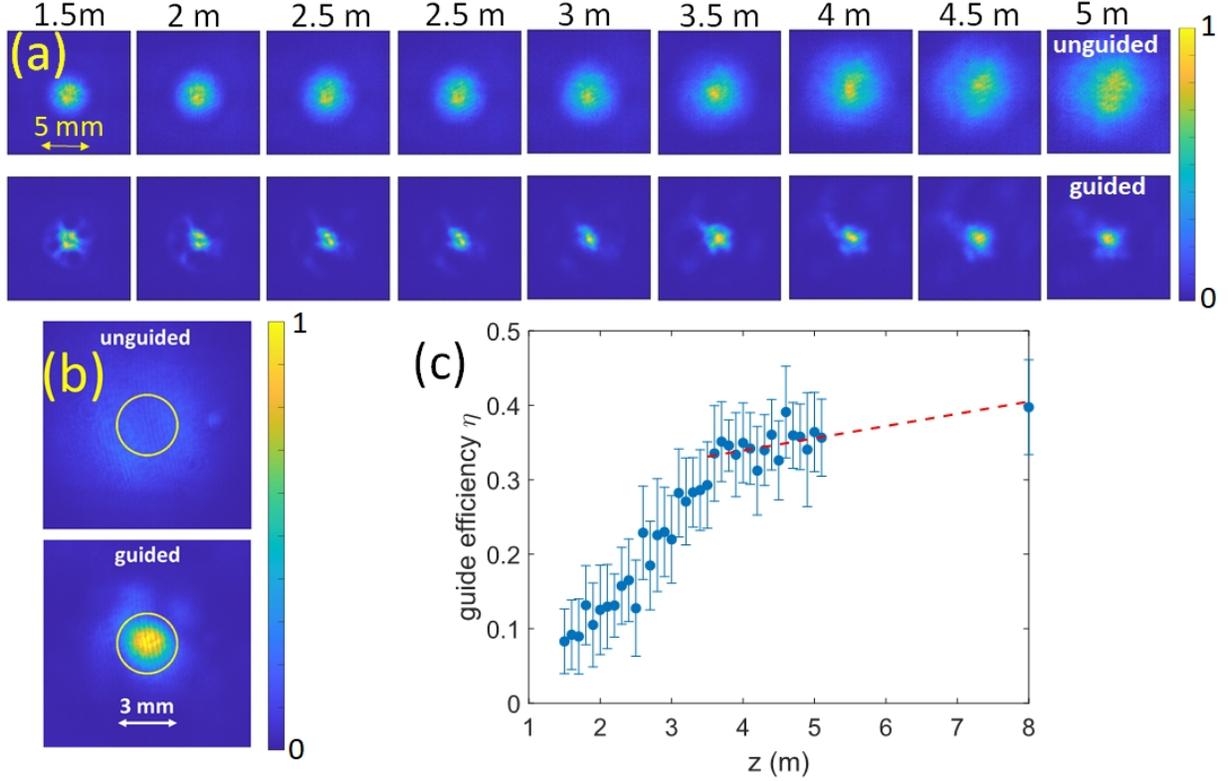

**Figure 6.** Optical guiding of a 1 mJ, 7 ns, λ=532 nm probe pulse injected at 800 μs delay into an air waveguide formed by a 80 mJ, 100fs $LG_{01}$ pulse with $d_{ring}$ = 3 mm. **(a)** 50 shot average images of unguided (top row) and guided (bottom row) mode vs. propagation distance. The guided mode diameter is ~2 mm. **(b)** Images at $z$ = 8 m of unguided and guided modes. The $d_{ring}$ = 3 mm diameter circles contain the guided and unguided energy used in the definition of guiding efficiency. **(c)** Guiding efficiency $\eta = (E_g - E_{ug})/(E_{tot} - E_{ug})$ vs. propagation distance, measured with helium cell. Each point is a 50 shot average, with the bars showing the ± standard deviation. The gap between points at $z$ = 5.3 m and $z$ = 8 m is due to helium cell travel constrained by our optical table arrangement. The dashed red line is a linear fit to the points > 3.5 m to guide the eye.

Results from a smaller diameter air waveguide are shown in Fig. 6. Here the waveguide generator was a 70 mJ, 100 fs $LG_{01}$ pulse with $d_{ring}$ = 3 mm ($w_0$ = 2.1 mm). The smaller ~2.5 mm guided mode diameter is immediately apparent in Fig. 6(a). Because the probe laser geometry is the same as for Fig. 5, the guiding efficiency, plotted in Fig. 6(b) and peaking at ~40%, is smaller consistent with the smaller waveguide numerical aperture.

Linear guiding of the diverging $f/950$ $\lambda$ = 532 nm probe pulse in our 8 m, $d_{ring}$ = 4.5 mm air waveguide is simulated using the beam propagation method (BPM) [37], which computes propagation assuming a time-independent laser field. This is appropriate for our very long-lived air waveguides. Figure 7(a) shows the cross sections of two air waveguides at 800 μs delay: one with 15 uniformly spaced density holes, and the other with 8 density holes and two azimuthal gaps. The peak hole depth variation along the guide plotted in Fig. 7(b), scaled using the measurement of Fig. 3(b). The guided modes for the 15-hole case are shown in Fig. 7(c), and clearly resemble the experimental images of Fig. 5, with similar rings and azimuthal modulations indicative of multimode guiding in an azimuthally modulated guide. The guiding efficiency $\eta$ is plotted in Fig. 7(d) for the waveguides of Fig. 7(a). For the 15-hole guide, the efficiency quickly rises from a minimum of ~60% to ~90% and stays at that level, higher than our peak experimental efficiency of ~60% from Fig. 5(d). An explanation for this difference is nonuniform azimuthal filament coverage for some sections of the waveguide in the experiment. This is borne out by the simulations: for the nonuniform 8-hole case, the peak guiding efficiency is significantly reduced. Future experiments will be dedicated to optimizing guiding efficiency by improving the azimuthal uniformity of filament coverage.

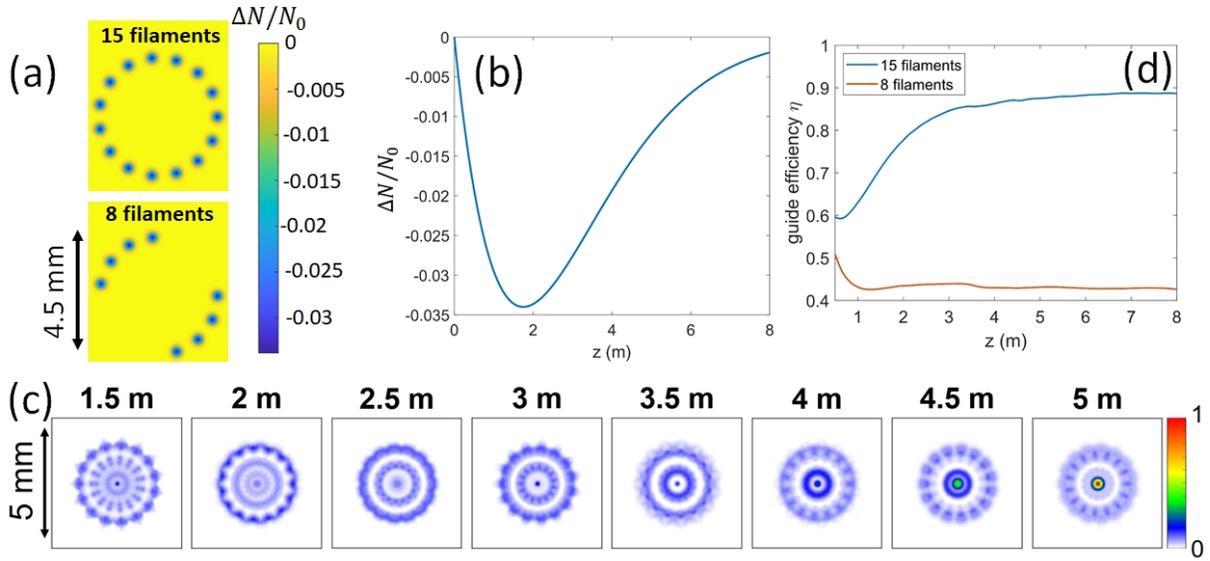

**Figure 7.** Beam propagation method (BPM) [37] simulation of linear λ=532 nm probe injection and propagation in an 8 m long air waveguide formed by a ring of filament-induced density holes. **(a)** $d_{ring} = 4.5$ mm diameter density hole array at 800 μs delay. Peak hole depth is $|\Delta N|_{max}/N_0 = 0.034$. Top panel: 15 filaments. Bottom panel: 8 filaments **(b)** Peak hole depth vs. axial location in guide, scaled from measurements in Fig. 3(b). **(c)** Guided mode profile vs distance in waveguide for 15-filament guide. The guide is highly multimodal, with $V^2/2 \sim 10^3$ modes. **(d)** Simulated guide efficiency $\eta$ (Eq. (2)) for 15-filament guide and 8-filament guide.

### (b) 50 m range air waveguiding experiments

Based on the detailed in-lab investigation described in the prior section, we next demonstrated air waveguiding over ~50 m in the hallway adjacent to the lab, using the setup of Fig. 2(b). To extend $LG_{01}$ filamentation over this longer range, we used $d_{ring} = 5.6$ mm ($z_0 \sim 60$ m), $E_{LG} = 120$ mJ (keeping the initial laser fluence $F_0$ nearly constant), and $\tau = 300$ fs (see Fig. 3(a) showing longer filaments with longer pulses). In the absence of the helium cell in the hallway, burn paper was used as a spatial profile diagnostic of the $LG_{01}$ filaments. Figure 4(d) shows a sequence of burns up to ~42 m, where reasonable azimuthal coverage is seen, with the number of filaments decreasing at the farthest positions owing to a decrease with propagation in the local fluence in the $LG_{01}$ ring. As in the in-lab experiments, the injected probe was a diverging $f/950$, 1 mJ, λ=532 nm pulse. Guided modes were imaged (through a 532 nm filter) from an alumina ceramic scattering screen variably placed along the propagation range. This method has lower fidelity than the direct imaging through the helium cell, but it allowed capture of the full beam without placing easily damaged optics in the path of the intense filaments.

Given the increased size of the $LG_{01}$ ring and the expected longer waveguide formation timescale and lifetime, we first measured guide efficiency $\eta$ (at $z = 42$ m) vs. probe pulse injection delay, as plotted in Fig. 8(a). An injection delay $\sim 2-5$ ms yields the maximum guiding efficiency $\eta \sim 15-20\%$. Efficiency vs propagation distance for a 5 ms injection delay is plotted in Fig. 8(b). The increase and then decrease of $\eta$ with delay in Fig. 8(a) appears as increasing and then decreasing mode confinement, as shown in the sequence of guided beam images in Fig. 8(c), measured at $z = 42$ m. While there is still noticeable beam confinement up to 20 ms delay, by 30 ms the guided and unguided modes are almost indistinguishable owing to thermal dissipation of the guiding structure.

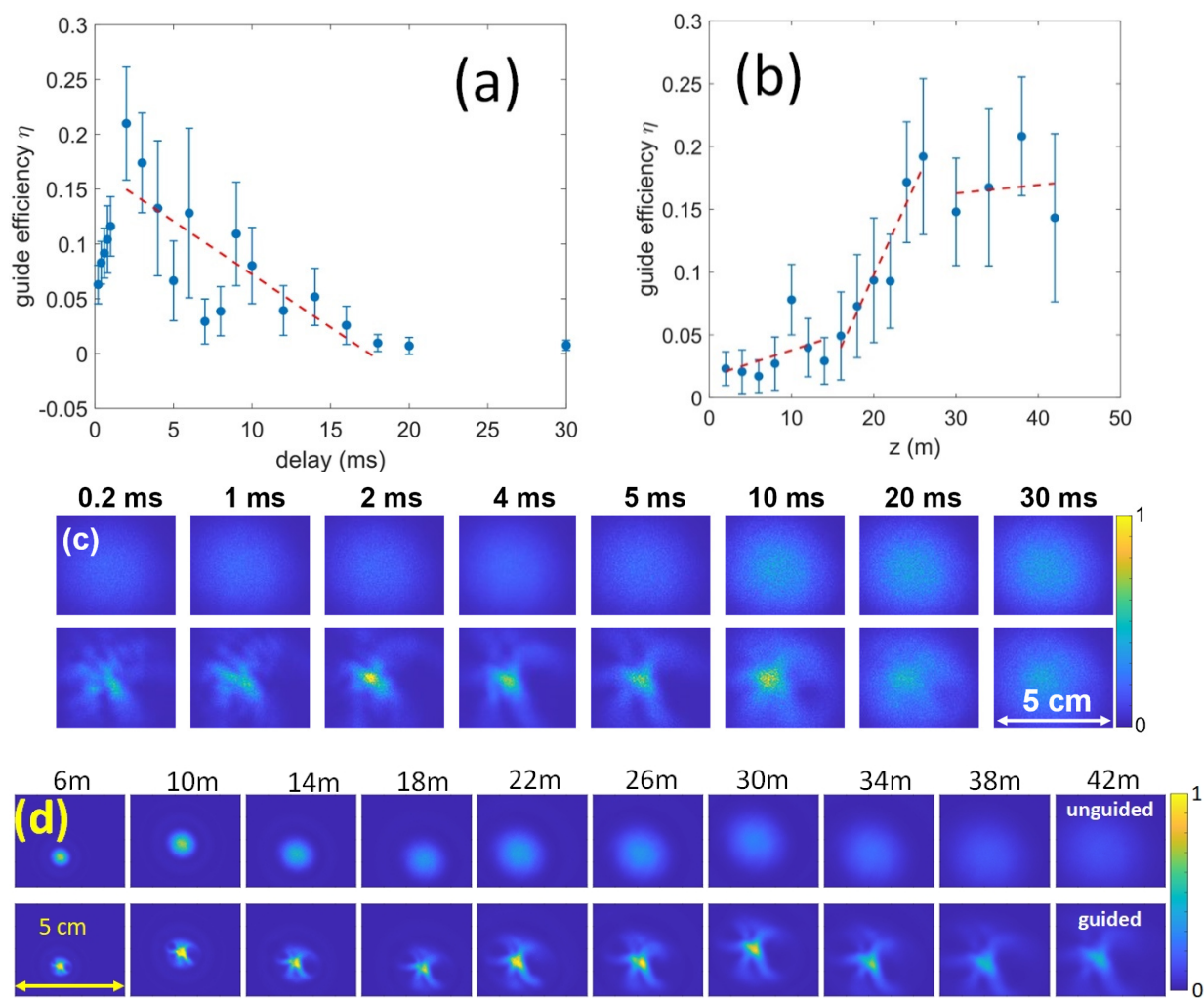

**Figure 8.** Results from air waveguiding in 50 m range. **(a)** Guiding efficiency $\eta$ vs. probe pulse injection delay, measured at $z = 42$ m. **(b)** Guiding efficiency $\eta$ vs. $z$ along waveguide. Each point in (a) and (b) is a 100 shot average, and the error bars are the ± standard deviation. **(c)** Guided probe modes measured at $z = 42$ m vs. injection delay. **(d)** 100 shot average images of unguided (top row) and guided (bottom row) mode vs. propagation distance.

A comparison of guided and unguided modes as a function of longitudinal position is shown in Fig. 8(d). As in the in-lab experiments, this is a highly multimodal beam. BPM simulations predict efficiency $\eta \sim 50\%$, here for 20 density holes at 5 ms delay on a ring of diameter $d_{ring} = 5.6$ mm, significantly higher than the experiment. As with the 8 m guide experiments and simulations, part of the discrepancy between simulated and measured efficiency is attributable to nonuniform and sparser azimuthal density hole coverage. This is evidenced by the burn patterns at longer distances in Fig. 4(d) and the asymmetric imprint of the density holes on the guided beam edge in Fig. 8(d). An additional important factor, especially for the ~50 m propagation experiments, is the probe beam pointing wander of ~200 mrad. Simulations show that ~200 mrad off-axis injection, into a guide with uniformly distributed density holes, decreases the efficiency to $\eta \sim 25\%$.

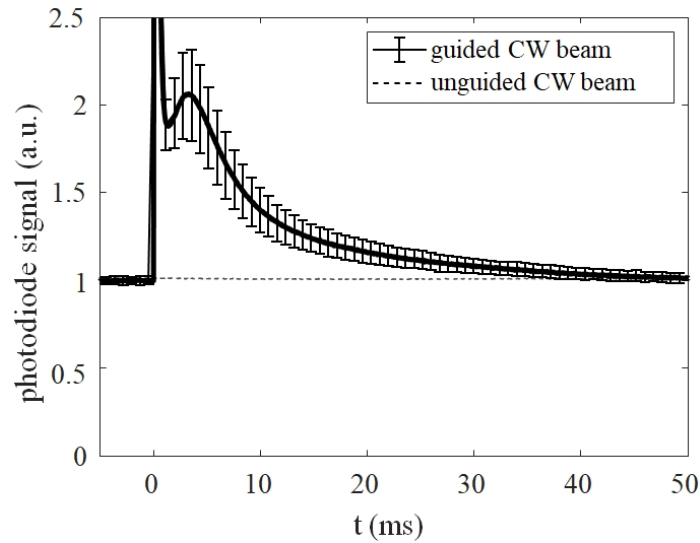

**Figure 9.** Measurement of guiding of a CW probe beam in a ~45 m waveguide. A 100 mW, $\lambda$=532 nm laser diode was injected into the waveguide through the down-collimation telescope mirror (see Fig. 2(b)), into the waveguide, and then collected by an integrating sphere at $z = 45$ m. The waveguide lifetime is directly read off the signal to be tens of milliseconds. Solid curve: 100 shot average. Error bars: ± standard deviation. The first peak is filament-generated supercontinuum collected by the integrating sphere and the second peak reflects maximum guiding efficiency at $\sim 3 - 5$ ms delay.

Finally, as a striking demonstration of the potential quasi-steady state operation of femtosecond filament-induced air waveguides, we injected a probe beam from a continuous wave (CW) 100 mW, $\lambda$=532 nm laser diode into our long hallway waveguide. The waveguide parameters were the same as for Fig. 8. The probe beam was injected along the green beam path shown in Fig. 2(b) and was collected at $z = $ ~45 m by integrating sphere, with a 1.3-cm entrance aperture slightly exceeding the mode size of Fig. 8. The goal was to directly measure the guiding lifetime of the waveguide. Figure 9 plots the average of 100 diode signal traces of the guided beam collected by the integrating sphere, with ± standard deviation bars overlaid on the curve. This shows that measurable guiding over 45 m persists to ~20 ms, declining to a negligible level at >30 ms owing to thermal diffusion of the waveguide. This is consistent with the ≥ 20 ms guided mode images of Fig. 8(c). The initial temporal spike in the diode signal is from $LG_{01}$ filament-induced supercontinuum emission that transmits through the 532nm interference filter in front of the integrating sphere. The secondary peak at ~3-5 ms follows the maximum guiding efficiency, corresponding to optimal waveguide cladding performance from diffusive merging of density holes; this is consistent with the peak in Fig. 8(a). These results strongly suggest that for similar guide parameters as in Figs. 8 and 9, a $> \sim 100$ Hz filament-forming laser could maintain a quasi-continuous air waveguide.

## V. CONCLUSIONS

We have demonstrated optical guiding of probe pulses up ~45 m, or 70 Rayleigh ranges, in air waveguides formed from the long-lived thermal response of air to filamenting ultrashort laser pulses. The guide lifetime is tens of milliseconds, showing that a quasi-steady state air waveguide could be supported by a $> \sim 100$ Hz repetition rate filament-forming laser. The guides are generated using a new method: multi-filamentation of Laguerre-Gaussian $LG_{01}$ "donut" modes, and are more than $60 \times$ longer than in our prior air waveguiding demonstration [19] which used a binary phase mask for filament formation. Our results pave the way for a wide range of applications requiring either projection or collection of optical signals in the atmosphere.

The long propagation range experiments, performed in the hallway adjacent to our laser lab, were preceded by more detailed measurements of waveguiding over 8 m (~13 Rayleigh ranges) in the lab. The in-lab experiments verified our new method for waveguide generation: the development of a thin ring of filaments by the $LG_{01}$ mode, which heat the air and form a cladding "moat" surrounding a core of unperturbed air. For the 8 m guides, maximum guiding efficiency was ~60%, and it was ~20% for the 45 m long waveguides. While lower than in the ideal case, these reduced efficiencies are likely due to nonuniform azimuthal coverage by the ring of filaments. This is directly borne out by simulations, which provide a path for straightforward efficiency improvements in future experiments.

While guiding in this experiment was demonstrated using a relatively weak probe laser pulse, air waveguides can support high peak laser powers and are especially suited to high average powers, limited in the first case by self-focusing and in the second case by thermal blooming. Using the scaling developed in ref. [19], the peak intensity $I_p$ limited by self-focusing is $I_p L_g < 2 \times 10^{14}$ W/cm, where $L_g$ is the guide length. Taking $L_g = 100$ m gives $I_p < 20$ GW/cm$^2$. For a ~5 mm diameter waveguide and a 10 ns pulse, the guided energy can be 40 J.

For quasi-CW guided beams, thermal blooming can limit the average laser power guided by spoiling the air waveguide structure. However, this will occur only at very large powers when sufficient laser energy is absorbed that the relative on-axis air temperature increase, $\delta_T = \Delta T/T_0$, dominates the waveguide core-cladding density contrast $\Delta N/N_0$. The relative temperature increase for average guided power $P_g$ over a duration $\Delta t$ is given by $P_g \Delta t / A = 1.5 \delta_T \alpha^{-1} p_0$, where $p_0$ is the ambient air pressure, $\alpha$ is the air absorption coefficient, and $A$ is the waveguide core cross-sectional area. For the ~8 m and ~50 m waveguides of this paper, a conservative estimate for the typical core-cladding density contrast is $\Delta N/N_0 \sim 0.01$, from inspection of Fig. 7(b). Using an absorption coefficient $\alpha = 2 \times 10^{-8}$ cm$^{-1}$, which accounts for molecular and aerosol contributions in realistic environments [38], a 5 mm waveguide core diameter, and a quasi-CW burst duration of 5 ms, gives an estimated thermal-blooming- limited guided energy of $P_g \Delta t \sim 15$ kJ and peak average power limit $P_g \sim 3$ MW.

For applications such as distant projection of high average powers or remote collection of optical signals [39], kilometer-length and longer air waveguides may be of interest. The laser requirements for their generation can be obtained from scaling the results of our experiments. Because filamentation occurs over approximately a Rayleigh range of the driving beam, an $LG_{01}$ beam of diameter $d_{beam} \sim 2w_0 = 3.2$ cm and sufficient power will generate filaments over ~1 km.

Each filament is much shorter than this, owing to the interaction of the filament with its local laser energy "reservoir", with some filaments decaying and others starting at multiple locations along the propagation path, but remaining, on average, uniformly spaced on the ring. Based on our ~50 m experiments, which used 120 mJ, 300 fs LG$_{01}$ pulses with $d_{ring} = 0.56$ cm ($d_{beam} \sim 2w_0 = 0.79$ cm), a 1 km air waveguide using $d_{beam} = 3.2$ cm would need $\sim(3.2/0.79)^2 \sim 16 \times$ more energy, or ~2 J, to maintain the fluence and keep the spacing between filaments on the ring approximately the same. Our prior measurements [18] have established average single filament energy deposition of $\sim 0.25 - 0.5$ µJ/cm in the extended propagation range of a filament. Therefore, for each density hole generated by filaments on the ring, an average $\sim 0.025 - 0.05$ J is needed over 1 km. A 2 J LG$_{01}$ pulse could then support ring coverage of $\sim 40 - 80$ filaments to form a waveguide cladding over ~ 1 km.

## ACKNOWLEDGEMENTS

A. Goffin and I. Larkin contributed equally to this work. This work is supported by the Office of Naval Research (N00014-17-1-2705 and N00014-20-1-2233), the Air Force Office of Scientific Research and the JTO (FA9550-16-1-0121, FA9550-16-1-0284, and FA9550-21-1-0405), the Army Research Lab (W911NF1620233) and the Army Research Office (W911NF-14-1-0372).

## APPENDIX A: NONLINEAR PROPAGATION SIMULATIONS

The 3D+1 simulations of LG$_{01}$ pulse propagation were performed using our UPPE (unidirectional pulse propagation equation) [40] implementation called YAPPE (yet another pulse propagation effort). UPPE is a system of ordinary differential equations of the form

$$\frac{\partial}{\partial z} A_{k_x,k_y}(\omega, z) = i2\pi Q_{k_x,k_y}(\omega) P_{k_x,k_y}(\omega, z) e^{-i\left(k_z - \frac{\omega}{v_g(\omega)}\right)z} . \tag{A.1}$$

In Eq. (A.1), $A_{k_x,k_y}(\omega, z)$ is the 3D inverse Fourier transform of the spacetime auxiliary field $A_{k_x,k_y}(\omega, z) = \mathcal{F}^{-1}_{x,y,\xi}\{E(x, y, \xi, z)e^{-ik_z\Delta z}\}$, where $\xi = t - z/v_g(\omega)$ is time in the pulse frame of reference and $\Delta z$ is the simulation step size. The transverse spatial frequencies $(k_x, k_y)$ index a system of ordinary differential equations which are numerically solved. $P_{k_x,k_y}(\omega, z)$ is the nonlinear polarization of the medium, including Kerr self-focusing, rotational nonlinearities [41], ionization dynamics [42], and a non-dispersive plasma response. The other variables are defined as follows: $\omega$ is the angular frequency, $v_g(\omega)$ is the group velocity of the medium as a function of frequency, $k_z = ((\omega/v_g(\omega))^2 - (k_x^2 + k_y^2))^{1/2}$ is the longitudinal spatial frequency, and $Q_{k_x,k_y}(\omega) = \omega/ck_z$. To recover the field in the spacetime domain, the auxiliary field is converted back to the electric field $E_{k_x,k_y}(\omega, z) = A_{k_x,k_y}(\omega, z)e^{ik_z\Delta z}$, and a 3D Fourier transform is performed on the electric field, $E(x, y, \xi, z) = \mathcal{F}_{k_x,k_y,\omega}\{E_{k_x,k_y}(\omega, z)\}$.